\documentclass[aps,showpacs,superscriptaddress,groupedaddress, twocolumn]{revtex4-1} 

\pdfoutput=1

\usepackage{amsmath}
\usepackage{amssymb}

\usepackage{graphicx}
\usepackage{subfigure}
\usepackage{longtable}

\usepackage{dcolumn}
\usepackage{bm}
\usepackage{array}
\usepackage{color}

\usepackage{longtable}%tables

\newtheorem{thm}{Theorem}%theorem

\graphicspath{{../1figures/}}

\begin{document}

%\begin{CJK*}{GBK}{song}

\title{A residual-based message passing algorithm for constraint satisfaction problems}

\author{Chun-Yan Zhao$^1$}
 \email{Corresponding author: zhaocy@usst.edu.cn}
 
\author{Yan-Rong Fu$^1$}
  
\author{Jin-Hua Zhao$^{2,3}$}
 \email{Corresponding author: zhaojh@m.scnu.edu.cn}
 
\affiliation{
$^1$College of Science,
University of Shanghai for Science and Technology,
Shanghai 200093, China}

\affiliation{
$^2$Guangdong Provincial Key Laboratory of Nuclear Science,
Institute of Quantum Matter,
South China Normal University,
Guangzhou 510006, China}

\affiliation{
$^3$Guangdong-Hong Kong Joint Laboratory of Quantum Matter,
Southern Nuclear Science Computing Center,
South China Normal University,
Guangzhou 510006, China}

\date{\today}

%arXiv version: 02.14.2022

\begin{abstract}
Message passing algorithms, whose iterative nature captures well complicated interactions among interconnected variables in complex systems and extracts information from the fixed point of iterated messages, provide a powerful toolkit in tackling hard computational tasks in optimization, inference, and learning problems. In the context of constraint satisfaction problems (CSPs), when a control parameter (such as constraint density) is tuned, multiple threshold phenomena emerge, signaling fundamental structural transitions in their solution space. Finding solutions around these transition points is exceedingly challenging for algorithm design, where message passing algorithms suffer from a large message fluctuation far from convergence. Here we introduce a residual-based updating step into message passing algorithms, in which messages varying large between consecutive steps are given high priority in the updating process. For the specific example of model RB, a typical prototype of random CSPs with growing domains, we show that our algorithm improves the convergence of message updating and increases the success probability in finding solutions around the satisfiability threshold with a low computational cost. Our approach to message passing algorithms should be of value for exploring their power in developing algorithms to find ground-state solutions and understand the detailed structure of solution space of hard optimization problems.
\end{abstract}

\maketitle

\section{Introduction}

The difficulty in understanding complex systems manifests not only in finding a proper description of their interaction topology,
but also in quantifying the collective and cooperative behaviors among their constituents at multiple scales \cite{Newman-AIP-2012}. Message passing algorithms prove to be an efficient tool in tackling hard computational problems formulated in graphical models \cite{Koller.Friedman-2009}, which is a simple description of complex interacted systems.
In a typical algorithmic procedure, one defines messages on the connections of basic units in a graphical representation, derives iterative equations of messages to account for the propagation of correlation among variables,
and extracts information on problem solutions through messages after a sufficiently large number of updating steps. Among the main roots of the message passing algorithms are coding theory
in information theory research \cite{Richardson.Urbanke-2008} and statistical physics of disordered systems
\cite{Mezard.Montanari-2009}. Message passing algorithms have been heavily explored in optimization problems
\cite{Garey.Johnson-1979,Zhou-2015}, inference problems \cite{Zdeborova.Krzakala-AdvInPhys-2016},
learning problems \cite{Braunstein.Zecchina-PRL-2006,Mezard-PRE-2017}, and so on.
Different frameworks of message passing algorithms
are based on mean-field theory at a certain level of approximation, whose typical examples encompass the belief propagation (BP) algorithm \cite{Mezard.Parisi-EPJB-2001}, the survey propagation algorithm \cite{Mezard.Parisi.Zecchina-Science-2002}, the cluster variational algorithm \cite{Kikuchi-PRB-1951,Pelizzola-JPhysA-2005}, and so on.

Constraint satisfaction problems (CSPs)
\cite{Garey.Johnson-1979} are among the most typical computational tasks in complex systems research. A typical instance of CSPs can be formulated as a bipartite graph with two types of nodes denoting $m$ constraints and $n$ variables respectively and edges only between constraints and variables. In a random setting, edges can be established from a null graph with only nodes
by connecting a given constraint node with multiple variable nodes chosen among the variable node set with a uniform probability.
A basic computational question for CSPs is to find the existence and the number of assignments to variables which satisfy all the constraints in a given instance. CSPs are heavily discussed in computational complexity theory in theoretical computer science \cite{Garey.Johnson-1979}, as many prototype NP-hard problems can be formulated as CSPs.
In the language of statistical physics, satisfiability and optimization problems can be understood as the energy landscapes of configurations of a physical system with continuous or discrete states,
while their global optima correspond to the ground-state energy of landscapes.
Thus we can adopt statistical mechanical frameworks, such as the mean-field theory of spin glasses \cite{Mezard.Montanari-2009},
to estimate the energy and the entropy of their ground-state solutions. Typical CSPs studied with the spin glass theory are the random $k$-satisfiability ($k$-SAT) problem \cite{Krzakala.etal-PNAS-2007,Montanari.RicciTersenghi.Semerjian-JStatMech-2008}, the minimum vertex cover problem \cite{Weigt.Hartmann-PRL-2000,Weight.Zhou-PRE-2006}, and the $q$-coloring problem \cite{Zdeborova.Krzakala-PRE-2007}. These problems focus on random CSPs with \textsl{fixed} domains, whose size of variables is independent of the variable number $n$ (that is, 2, 2, and $q$ as given integers, respectively). Yet many practical problems, such as the Latin square problem, the $n$-queens problem and so on, can be transformed into random CSPs with \textsl{growing} domains. Among the models for these CSPs, model RB (revised B) proposed in \cite{Xu-JAIR-2000} is a popular one, whose domain size grows polynomially with the variable number $n$.

Spin glass mean-field theory on satisfiability problems shows that fundamental structural transitions exist in the geometric organization of ground-state configurations. Taking the random $k$-SAT ($k \geqslant 3$) problem \cite{Krzakala.etal-PNAS-2007} as an example, with an increase in constraint densities (ratio of constraint number to variable number), there are clustering, condensation, and SAT/UNSAT phase transitions separated by multiple thresholds, where drastic changes in the numbers and sizes of the clusters of ground-state configurations take place. Furthermore, there is a deep connection between these solution space phase transitions and the algorithmic performance in finding solutions. Specifically, around these thresholds, messages in mean-field theory often show large fluctuations, rendering impossible the convergence of messages from which a typical message passing algorithm can extract information. Meanwhile, each mean-field theory has an approximation at some level on the organization of the solution space. For example, for the BP algorithm \cite{Mezard.Parisi-EPJB-2001}, solutions are assumed to be organized in a single cluster, or a pure state. When a level of approximation fails, there are often two choices: whether we generalize the current mean-field assumption to a theoretical framework with a more complicated and computationally demanding form, or we introduce modifications at an algorithmic level to achieve better performance by leveraging the existing form of message passing algorithm.

Our focus in this paper is to suppress the fluctuation of iterative messages around thresholds, a critical problem for message-passing-based algorithms, thus to improve the performance of existing BP algorithm for solving CSPs at the algorithmic level.
Our main contribution here is to define residuals in the message updating process, based on which the existing BP algorithm is further modified.
We show that, for a specific CSP model RB, compared with its usual BP algorithm,
our modification of the algorithm can significantly improve the convergence of the updated messages and increase the probability of finding ground-state solutions around the satisfiability thresholds at a lower computational cost.

The layout of the paper is as follows. In Section $2$, we briefly introduce the concept and some known results of model RB, which is a typical CSP with growing domains. Section $3$ presents the maximal residual BP (MRBP) algorithm, which modifies the current BP algorithm based on residuals in the process of updating messages. In Section $4$, we illustrate and analyze the numerical results of our algorithm. In Section $5$, we conclude the paper with some discussions.

\section{Model RB: an example of constraint satisfaction problems}

Over the past few decades, studies on CSPs mainly focus on the initial standard CSP models, named A, B, C and D \cite{Smith-AI-1996,Gent-C-2001,Prosser-AI-1996}. However, it has been proved by Achlioptas et al. that the random instances generated by the four standard models have trivial asymptotic insolubility with an increase in the problem size \cite{Achlioptas-C-2001}. Therefore, in order to overcome this defect, various improved models from different perspectives in numerous efforts have been proposed successively \cite{Xu-JAIR-2000,Molloy-SIAMJC-2003,Frieze-RSA-2006,Smith-TCS-2001,Gao-JAIR-2007}. Model RB, proposed by Xu and Li to avoid the trivial unsatisfiability of model B \cite{Xu-JAIR-2000}, is a modification on the domain size of variables and the number of constraints of model B without special restrictions on the constraint structure.

Model RB consists of a set of $n$ variables $X=\{x_1, x_2, \ldots, x_n\}$ and a set of $m$ constraints $C = \{C_1, C_2, \ldots, C_m\}$ with $m = r n \ln n$, where $r>0$ is a constant. Each variable $x_i$ ($i=1,2,\ldots,n$)  has a nonempty domain $D=\{1, 2, \ldots, d_n\}$ of $d_n=n^{\alpha}$ ($\alpha>0$ is a constant) possible values. Each constraint $C_a$ ($a=1,2,\ldots,m$) involves $k$ ($\geqslant 2$) different variables, and there is a set of incompatible assignments $Q_a\subset D^k$ which specifies the disallowable combinations of values for the $k$ variables.
More precisely, we generate a random instance of model RB with the following two steps.
\begin{description}
\item[Step 1.]
To construct a single constraint, we randomly select $k$ ($\geqslant 2$) distinct variables out of the $n$ variables, and then randomly select exactly $pd_n^k$ ($p$ is the constraint tightness) different incompatible assignments out of $d_n^k$ possible ones as the set $Q_a$ to restrict the values of these $k$ variables.
\item[Step 2.]
We repeat the above step to obtain $m=r n \ln n$ independent constraints and their corresponding sets $Q_a$ $(a = 1, 2, \cdots, m)$.

\end{description}
A random instance of model RB is simply the conjunction of the above $m$ randomly selected constraints.

Let $\Pr(\rm SAT)$ be the probability that a random instance is satisfied.
It is shown in Ref.~\cite{Xu-JAIR-2000} that

\begin{thm}
Let $p_s=1-{\rm e}^{-\frac{\alpha}{r}}$. If $\alpha>\frac{1}{k}$, $r>0$ are two constants and $k {\rm e}^{-\frac{\alpha}{r}} \geqslant1$, then
\begin{equation}
\lim_{n\rightarrow\infty}\Pr(\rm SAT)=\left\{
                                    \begin{array}{ll}
                                      1, & \hbox{if $p<p_s$;} \\
                                      0, & \hbox{if $p>p_s$.}
                                    \end{array}
                                  \right.
\end{equation}
\label{thm1}
\end{thm}
\begin{thm}
Let $r_s=-\frac{\alpha}{\ln(1-p)}$. If $\alpha>\frac{1}{k}$, $0<p<1$ are two constants and $k \geqslant \frac{1}{1-p}$, then
\begin{equation}
\lim_{n\rightarrow\infty}\Pr(\rm SAT)=\left\{
                                    \begin{array}{ll}
                                      1, & \hbox{if $r<r_s$;} \\
                                      0, & \hbox{if $r>r_s$.}
                                    \end{array}
                                  \right.
\end{equation}
\end{thm}
The two theorems show that model RB exhibits a sharp threshold of satisfiability at the critical values $p_s$ and $r_s$, as $p$ and $r$ are two control parameters of the problem.  Nevertheless, we cannot rigorously determine the exact value of satisfiability threshold for some random CSPs with fixed domains \cite{Achlioptas-Nature-2005,Mertens-RSA-2006}. It has been theoretically proved that almost all instances of model RB have no tree-like resolution proofs of less than exponential size \cite{Xu-TCS-2006}. In other words, these random instances of model RB in the transition region are extremely difficult to solve. Numerical results, using complete and incomplete search algorithms, have confirmed the exact satisfiability thresholds and the hardness of forced and unforced satisfiable instances \cite{Xu-AI-2007}. Although model RB shows an asymptotic phase transition phenomenon, it is still very challenging to find solutions of a random instance, and the size of the problem has been restricted to $n \sim100$ \cite{Cai-AI-2011}. Therefore, benchmarks based on model RB are widely used in various kinds of algorithm competitions such as CSP, SAT, pseudo-Boolean and so on, and these results have verified the intrinsic computational hardness of these benchmarks.

Both analytical and algorithmic studies have been carried out on model RB. On the analytical side, Fan and Shen defined a large class of random CSP models $d$-$k$-CSP which unified several related models including model RB \cite{Fan-AI-2012,Shen-JCO-2016}. Later, it was proved that the restriction on $k$ can be weaker, which can simplify the generation of random instances \cite{Zhou-JCO-2015}.
In order to study the transition behaviors of model RB due to finite-size effects, Zhao et al. use finite-size scaling analysis to bound the width of the transition region \cite{Zhao-IPL-2011}. On the algorithmic side, statistical physics concepts and methods, especially the replica method and the cavity method in spin glass mean-field theory, have played a very significant role in constructing solutions for random CSPs \cite{Olivier-TCS-2001,Marino-NC-2016}.  Zhao et al. proposed two different types of message passing algorithms guided by BP to solve random CSPs with large domains like model RB \cite{Zhao-JSTAT-2011}. Subsequently,  Zhao et al. proposed a reinforced BP algorithm, which can effectively force the BP equations to converge to a solution of model RB by introducing an external field with a certain probability \cite{Zhao-PRE-2012}. Moreover, it was shown that the solutions of model RB are grouped in disconnected clusters before the theoretical satisfiability phase transition threshold \cite{Zhao-PRE-2012}. Recently, Zhao et al. proposed three kinds of BP decimation algorithms, which improved the performance of BP algorithms by improving the updating method of BP iterative equation \cite{Zhao-JSTAT-2021}.

\section{Maximal residual belief propagation algorithm}

Here we propose the MRBP algorithm, which combines modifications in the updating process based on residuals with the usual BP algorithm.
We first lay down the general procedure of the BP algorithm and the relevant equations, then we define the residuals of messages in the updating process and present our algorithm in detail.

For $k \geqslant 2$, model RB is NP-complete. Therefore, we take the binary case ($k=2$) of model RB as the tested problem. Model RB with $k \geqslant 3$ can be reduced to the case of $k=2$ in polynomial time.

\subsection{Belief propagation algorithm}

\begin{figure}
\begin{center}
 \includegraphics[width = 0.70 \linewidth]{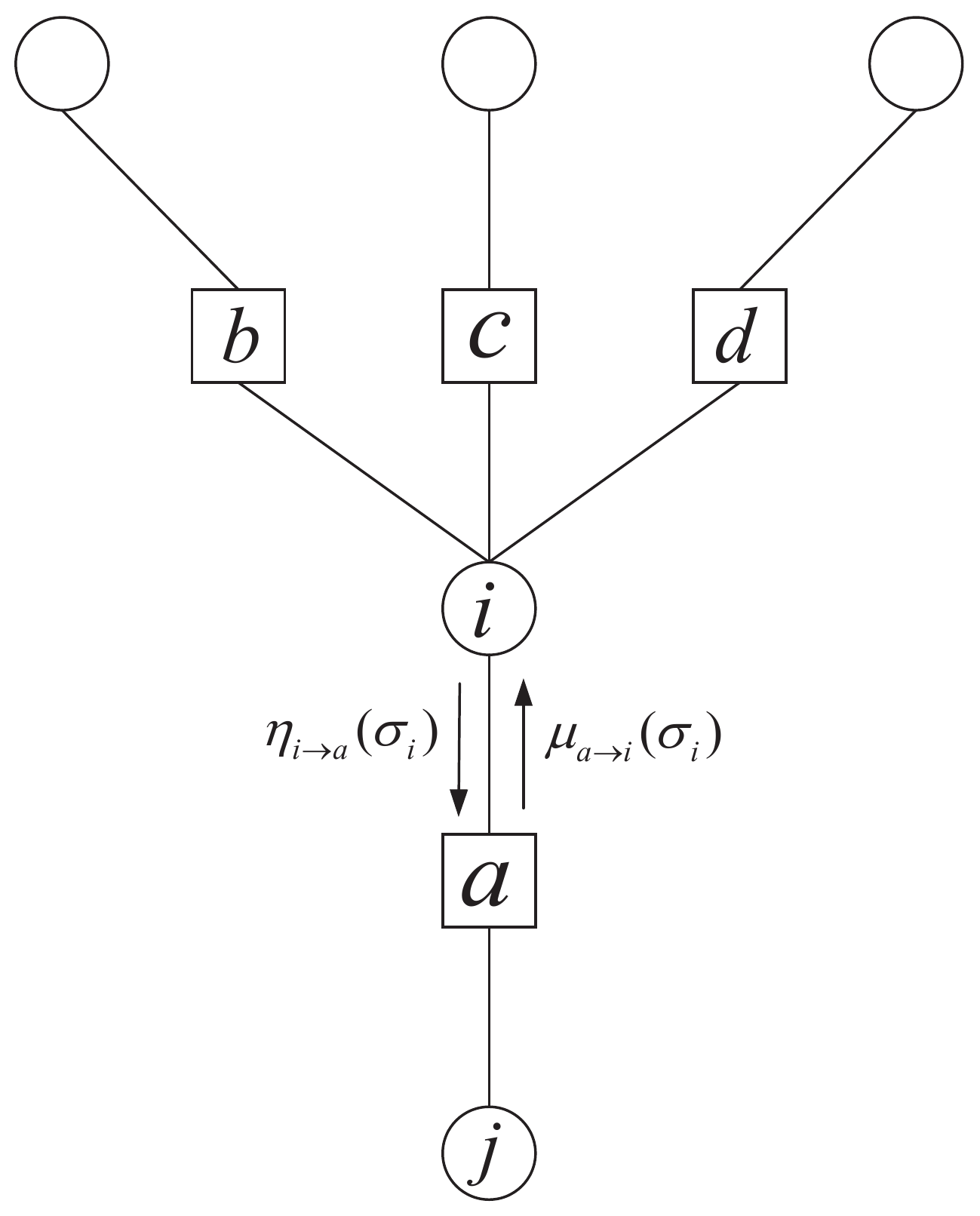} %0.80
\end{center}
\caption{
 \label{fig:fig1}
Part of a factor graph representing a random instance of binary model RB. Circles and squares denote the variable nodes and the constraint nodes, respectively. In the message passing algorithm of model RB, $\mu_{a\rightarrow i}(\sigma_i)$ and $\eta_{i\rightarrow a}(\sigma_i)$ are the messages passed between  $a$ and $i$ along the edge $(a,i)$ in opposite directions.}
\end{figure}

An instance of binary model RB admits a natural factor graph representation, as shown in figure~1. A factor graph is a bipartite undirected graph, in which one type of nodes represents the variables (denoted by $i, j, \cdots$) and the other type of nodes represents the constrains (denoted by $a, b, \cdots$). An edge $(a,i)$ connects a constraint node $a$ and a variable node $i$ if and only if $a$ contains $i$. The average degree of each variable node in the graph is $r\ln n$. Locally such a graph is tree-like: the typical size of a loop in the graph scales like $\ln n/\ln\ln n$ for large $n$.

An instance of model RB can be treated as a statistical mechanical system with interacting variables with discrete states and an interaction topology in the form of a factor graph.
In a random instance of binary model RB,
we let $\vec{\sigma}=(\sigma_1, \sigma_2, \ldots, \sigma_n)$ ($\sigma_i\in D$, $i=1,2, \cdots, n$) be an assignment to the $n$ variables.
We further let $\vec{\sigma_a}=(\sigma_{a1}, \sigma_{a2})$ be an assignment to the two variables involved in the constraint $a$ ($\in \{1, 2, \cdots, m\}$). We define the energy of the constraint $a$ as
$f_a(\vec{\sigma_a})=1$ if $\vec{\sigma_a}\in Q_a$ and $f_a(\vec{\sigma_a})=0$ if $\vec{\sigma_a}\notin Q_a$. The total energy of an assignment $\vec{\sigma}$, denoted as $E(\vec{\sigma})$, is the sum of energy terms on all the $m$ constraints, which is the number of unsatisfied constraints computed following
\begin{equation}
E(\vec{\sigma})=\sum_{a=1}^m f_a(\vec{\sigma_a}).
\label{eq3}
\end{equation}

Based on the locally tree-like property of the factor graph representation of model RB, we derive the framework of BP algorithm \cite{Mezard.Montanari-2009}. Associated with each edge $(a, i)$ there are two cavity messages $\mu_{a\rightarrow i}(\sigma_i)$ and $\eta_{i\rightarrow a}(\sigma_i)$ passing in opposite directions. See figure 1 for an example. Specifically, $\mu_{a\rightarrow i}(\sigma_i)$ is the probability that the constraint $a$ is satisfied if the variable $i$ takes the value $\sigma_i$, and $\eta_{i\rightarrow a}(\sigma_i)$ is the probability that the variable $i$ takes the value $\sigma_i$ in the absence of the constraint $a$. Let $\{\mu_{a\rightarrow i}^t(\sigma_i) \}$ and $\{ \eta_{i\rightarrow a}^t(\sigma_i) \}$ denote the set of messages passed along all the edges in the factor graph at time $t (= 0, 1, \cdots)$.
According to the cavity method in disordered systems of statistical physics
\cite{Mezard.Montanari-2009,Zhao-JSTAT-2011}, the self-consistent BP equations of cavity messages can be derived as
\begin{eqnarray}
\label{eq4}
\eta_{i\rightarrow a}^{t}(\sigma_i)
&=&\frac{1}{Z_{i\rightarrow a}}\prod_{b\in V(i)\backslash a}\mu_{b\rightarrow i}^t(\sigma_i), \\
\label{eq5}
\mu_{a\rightarrow i}^{t+1}(\sigma_i)
&=&\frac{1}{Z_{a\rightarrow i}}\sum_{j \in V(a)\backslash i, \sigma_j \in D}f_a(\vec{\sigma_a})\eta_{j\rightarrow a}^{t}(\sigma_j).
\end{eqnarray}
Here $Z_{i\rightarrow a}$ and $Z_{a\rightarrow i}$ are both normalization factors, $V(i)\backslash a$ the set of constraints adjacent to the variable $i$ removing $a$, and $V(a)\backslash i$ the set of variables adjacent to the constraint $a$ removing $i$.

In a typical procedure of the BP algorithm, messages on the edges in the factor graph of a given problem instance are uniformly initialized at random in $[0, 1]$.
A message iteration process is carried out,
in which messages are updated following the self-consistent equations (\ref{eq4}) and (\ref{eq5}). After a sufficiently long iterative process, the messages may converge to a fixed point. The convergence criterion can be measured by a precision parameter $\varepsilon$ ($ \ll 1$): when the maximal absolute difference between two consecutive steps of the messages on all edges of a factor graph is smaller than $\varepsilon$, we can say that the message updating process converges. With the fixed point of messages, statistical mechanical properties of model RB in an average sense can be extracted, such as energy and entropy of ground-state solutions, and so on.

To find a solution configuration  of a given problem instance,
we adopt the BP decimation (BPD) algorithm, which assigns values to variables one by one based on the marginal probability after message updating.
In order to find a solution configuration of the problem, we can assign the value of each variable based on its marginal probability distribution. This is the basic procedure of the BPD algorithm.
With the fixed point of messages
$\{\mu_{a\rightarrow i}^{\ast}(\sigma_i)\}$ for all edges $(a, i)$, the marginal probability of a variable $\sigma _{i}$ $(i = 1, 2, \cdots, n)$ is computed following
\begin{equation}
\mu_{i}(\sigma_i)
= \frac{\prod_{a\in V(i)}\mu_{a\rightarrow i}^{\ast}(\sigma_i)}{\sum_{\sigma_i\in D}\prod_{a\in V(i)}\mu_{a\rightarrow i}^{\ast}(\sigma_i)}.
\label{eq6}
\end{equation}
In a decimation step, the variable with the largest marginal probability is selected to be assigned to its corresponding most biased component. In the decimation procedure, we iteratively switch between a variable fixing process and a message updating process after a graph simplification process upon fixing variables.

\subsection{Maximal residual belief propagation algorithm}

\begin{figure}
\begin{center}
 \includegraphics[width = 0.99 \linewidth]{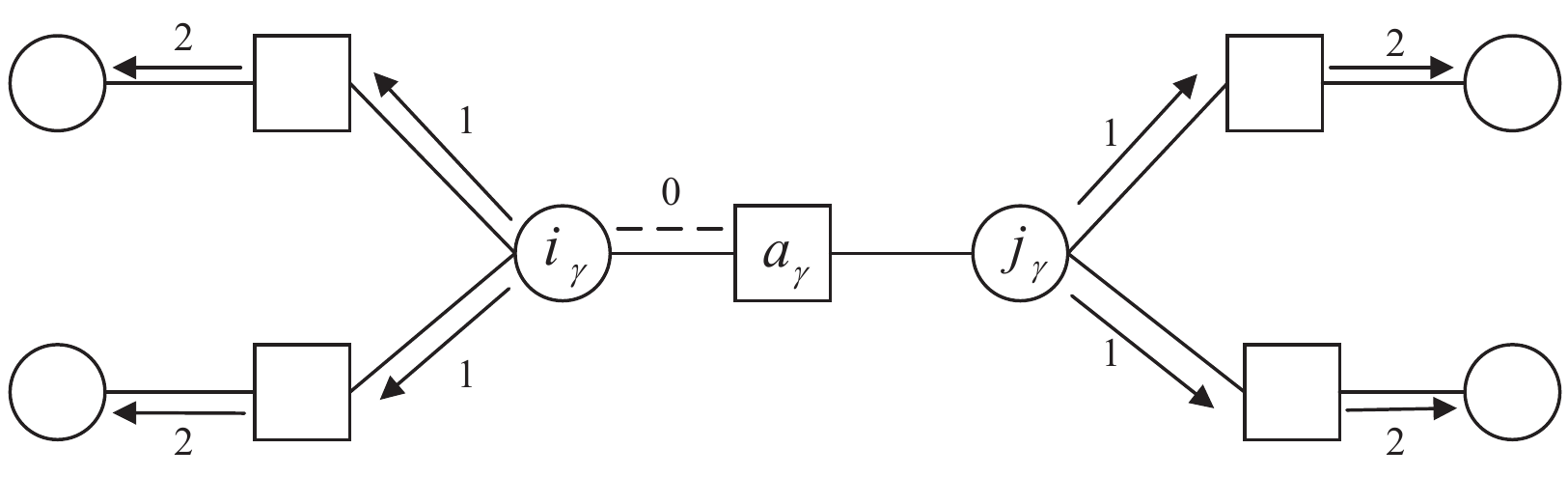} %0.99
\end{center}
\caption{
 \label{fig:fig2}
Local procedure of message updating based on maximal residuals on a factor graph.
The edge $(a_{\gamma},i_{\gamma})$ with the largest residual is selected for updating and then marked with $0$, and the numbers 1 and 2 represent the order of the message updating.}
\end{figure}
\begin{figure}
\begin{center}
 \includegraphics[width = 0.99 \linewidth]{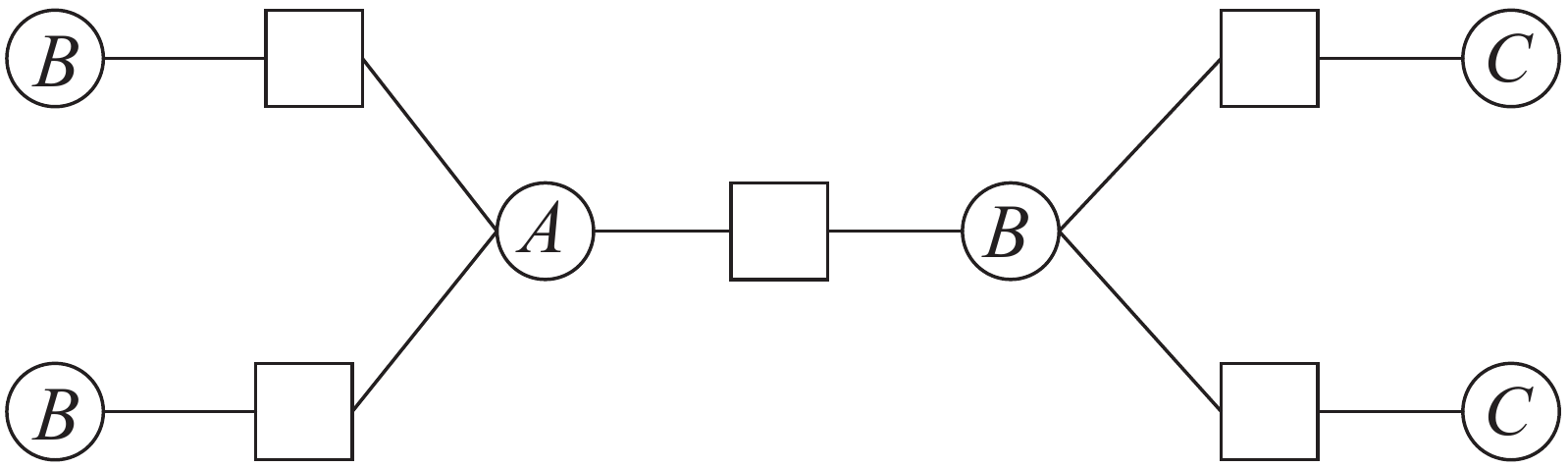} %0.99
\end{center}
\caption{
 \label{fig:fig3}
In the decimation process of the BPD algorithm, we classify the variables into three different types, indicated by A, B and C.}
\end{figure}

The convergence of messages in the BP iterative equations affects the performance of the algorithm. However, the BP equations usually do not converge when the control parameter approaches the theoretical satisfiability threshold.
The core idea of this work is the introduction of the residuals for messages.
We modify the message updating process,
as we dynamically update the passing messages between the constraints and the variables based on the maximal residual in BP iterative equations, to improve the convergence and the performance of BP algorithm.
The residuals of messages measure the relative difference sent from constraints to variables between
two consecutive steps during the iteration of messages.
If we consider all the messages in the BP algorithm as a time series are $\{ \mu_{a\rightarrow i}^t \}$ with $t = 0,1, \cdots$,
the residual on an edge $(a,i)$ at time $t$ is defined as
\begin{equation}
\gamma _{a \rightarrow i}^{t}
= \frac{| \mu_{a \rightarrow i}^t-\mu_{a\rightarrow i}^{t-1} |}{\mu_{a\rightarrow i}^t}.
\label{eq7}
\end{equation}
A large residual indicates a large fluctuation of the message $\mu_{a \rightarrow i}^t$,
thus the messages at these steps are far from the value after convergence of the whole message passing algorithm, if there is one.

Based on the residual on each edge at a time step,
the MRBP algorithm dynamically adjusts the order of updating messages according to the real-time status of the messages,
in which the messages connected to the edge with the largest residual are prioritized to update.
See figure 2 for an example.
At time $t$, we select the edge $(a_{\gamma},i_{\gamma})$ with the largest residual.
We first adopt equation $(4)$ to compute the messages sent from the variables $i_{\gamma}$ and $j_{\gamma}$
associated with the constraint node $a_{\gamma}$ to the neighboring constraint nodes connected to $i_{\gamma}$ and $j_{\gamma}$ except $a_{\gamma}$ (denoted by $V(i_{\gamma})\cup V(j_{\gamma})\setminus a_{\gamma}$),
as indexed by the number 1 in figure 2.
We need to reduce the greediness of the algorithm to achieve better performance. Hence, the updating messages are considered from the point of view of node $a_{\gamma}$ rather than edge $(a_{\gamma},i_{\gamma})$. Therefore, we use equation $(4)$ to calculate the messages sent from the nodes $i_{\gamma}$ and $j_{\gamma}$ connected to $a_{\gamma}$ to $V(i_{\gamma})\cup V(j_{\gamma})\setminus a_{\gamma}$ simultaneously.
Then, we follow equation $(5)$ to update the messages that the constraint nodes $V(i_{\gamma})\cup V(j_{\gamma})\setminus a_{\gamma}$
send to the variables connected to the constraints except $i_{\gamma}$ and $j_{\gamma}$,
as indexed by the number 2 in figure 2.
In order to avoid falling into the local optimum in the iteration process,
we mark the edge with the largest residual after the messages updating,
and the residuals on unmarked edges are recalculated.
The selection of the current edge with largest residual and the recalculation of residuals on unmarked edges
are carried out iteratively until all edges are marked to ensure that messages on all edges are updated during the BP iteration process.
In this way, the messages between constraints and variables are constantly updated till they converge to a fixed point or the maximal number of iterations is reached.

During the decimation process of the BPD algorithm, we iteratively fix variables with values until we have an assignment for all variables.
Once a variable in a factor graph is assigned with a certain value,
variables in the factor graph are divided into three different types, as illustrated in figure 3.

\begin{itemize}
  \item \textbf{Type A:} The variables that have been assigned with some values.
  \item \textbf{Type B:} The variables not assigned with values, yet connected to the same constraints with the assigned ones.
  \item \textbf{Type C:} Others.
\end{itemize}

The MRBP algorithm are detailed in the following pseudocode.

\noindent \rule[1pt]{86.0mm}{1pt} \\
{\bf MRBP algorithm} \\
\rule[1pt]{86.0mm}{1pt}

\noindent {\bf Input}: A factor graph of a random instance generated by model RB, maximal iterations $t_{\rm max}$ used in the subroutines MRBP-UPDATE and MRBP-UPDATE$^{\ast}$, and a precision parameter $\varepsilon$.

\noindent {\bf Output}: A solution or `UNCONVERGED'.

\begin{description}
  \item[Step 1] At $T=0$:
                \begin{description}
                    \item[1.1] Run the subroutine MRBP-UPDATE.
                    \item[1.2] For each variable $i$, compute the marginal probability $\mu_{i}(\sigma_i)$ by equation $(6)$.
                    \item[1.3] Select the variable $i^{\ast}$ with the highest marginal probability from the $n$ variables, and assign it to its most biased value $\sigma_{i^{\ast}}^{\ast}=\arg\max \mu_{i^{\ast}}(\sigma_{i^{\ast}})$.
                \end{description}
  \item[Step 2] From $T=1$ to $n-1$:
                \begin{description}
                    \item[2.1] Run the subroutine MRBP-UPDATE$^{\ast}$.
                    \item[2.2] For each free (unfixed) variable $i$, compute the marginal probability $\mu_{i}(\sigma_i)$ by equation $(6)$.
                    \item[2.3] Select the variable $i^{\ast}$ with the highest marginal probability from the $(n-T)$ free variables,
                    and assign it to its most biased value
                    $\sigma_{i^{\ast}}^{\ast}=\arg\max \mu_{i^{\ast}}(\sigma_{i^{\ast}})$.
                    \item[2.4] Check the energy function $E(\vec{\sigma})$ for the variables with assigned values according to equation $(3)$.
                    If $E(\vec{\sigma}) > 0$, break the loop and output `UNCONVERGED'.
                \end{description}
  \item[Step 3] If the assignment $\vec{\sigma}^{\ast}=(\sigma_1^{\ast}, \sigma_2^{\ast}, \ldots, \sigma_n^{\ast})$ of the $n$ variables satisfy the $m$ constraints simultaneously, output $\vec{\sigma}^{\ast}$ as a solution of the instance, otherwise output `UNCONVERGED'.
\end{description}
\rule[1pt]{86.0mm}{1pt}

The subroutine MRBP-UPDATE
is to update the messages on edges of a factor graph until a fixed point or a maximal number of iteration steps.

\noindent\rule[1pt]{86.0mm}{1pt}\\
Subroutine MRBP-UPDATE:
\begin{description}
\item[Step 1] At $t=0$: for each edge $(a, i)$, randomly initialize $\mu_{a\rightarrow i}^{0}(\sigma_{i})\in [0,1]$.
\item[Step 2] For the constraints from $a=1$ to $m$:
              \begin{description}
                \item[2.1] For each edge $(a,i)$ with $i\in V(a)$,
                use equation $(4)$ to obtain $\eta_{i\rightarrow a}^{0}(\sigma_i)$.
                In the case of $V(i)\backslash a=\phi$, set $\eta_{i\rightarrow a}^{0}(\sigma_i)=1/d_n$ for any $\sigma_i\in D$.
                \item[2.2] Use equation $(5)$ to calculate $\mu_{a\rightarrow i}^{1}(\sigma_i)$.
              \end{description}

\item[Step 3] From $t=1$ to $t_{\rm max}$:
              \begin{description}
                \item[3.1] For each edge $(a,i)$, calculate the residual $\gamma_{a\rightarrow i}^t$ using equation $(7)$.
                Select the edge $(a_\gamma,i_\gamma)$ with the largest residual among all edges and mark it.
                \item[3.2] For $i\in V(a_\gamma)$, use equation $(4)$ to calculate $\eta_{i\rightarrow a}^{t}(\sigma_i)$,
                where $a\in V(i)\backslash a_\gamma$. Update $\mu_{a\rightarrow j}^{t+1}(\sigma_j)$ using equation $(5)$, where $j\in V(a)\backslash i$.
                \item[3.3] For each edge $(a,i)$,
                if $|\mu_{a\rightarrow i}^{t+1}(\sigma_{i})-\mu_{a\rightarrow i}^{t}(\sigma_i)| <\varepsilon$
                holds, break the loop and set
                $\mu_{a\rightarrow i}^{\ast}(\sigma_i)=\mu_{a\rightarrow i}^t(\sigma_{i})$.
                Otherwise go to \textbf{3.1} until all edges are marked.
               \item[3.4] If $t=t_{\rm max}$, output `UNCONVERGED'.
              \end{description}
\end{description}
\rule[1pt]{86.0mm}{1pt}

The subroutine MRBP-UPDATE$^{\ast}$
is to obtain the fixed point of message updating in the decimation process,
during which some variables have already been assigned with certain values.

\noindent \rule[1pt]{86.0mm}{1pt}\\
Subroutine MRBP-UPDATE$^{\ast}$:
\begin{description}
\item[Step 1] At $t=0$:
    \begin{itemize}
        \item For variables of \textbf{Type A}: skip;
        \item For variables of \textbf{Type B}: if the assignment of the variable $i$ and the value of the fixed variable satisfy the corresponding constraint $a$, set $\mu_{a\rightarrow i}^{0}(\sigma_i)=1$; otherwise, set $\mu_{a\rightarrow i}^{0}(\sigma_i)=0$;
        \item For variables of \textbf{Type C}: uniformly initialize the values of $\mu_{a\rightarrow i}^{0}(\sigma_{i})\in [0,1]$ at random.
    \end{itemize}

\item[Step 2] For the constraints from $a=1$ to $m$:
              \begin{description}
                \item[2.1] For each edge $(a,i)$, where $i\in V(a)$.
                           \begin{itemize}
                             \item For variables of \textbf{Type A} and \textbf{Type B}: skip;
                             \item For variables of \textbf{Type C}: use equation $(4)$ to obtain $\eta_{i\rightarrow a}^{0}(\sigma_i)$.
                            In the case of $V(i)\backslash a=\phi$, set $\eta_{i\rightarrow a}^{0}(\sigma_i)=1/d_n$ for any $\sigma_i\in D$.
                             \end{itemize}
                \item[2.2] Update $\mu_{a\rightarrow i}^{1}(\sigma_i)$.
                           \begin{itemize}
                             \item For variables of \textbf{Type A}: skip;
                             \item For variables of \textbf{Type B}: let $\mu_{a\rightarrow i}^{1}(\sigma_i)=\mu_{a\rightarrow i}^{0}(\sigma_{i})$;
                             \item For variables of \textbf{Type C}: use equation $(5)$ to update $\mu_{a\rightarrow i}^{1}(\sigma_i)$.
                           \end{itemize}
              \end{description}

\item[Step 3] From $t=1$ to $t_{\rm max}$:
              \begin{description}
                \item[3.1] Compute the residual $\gamma_{a\rightarrow i}^t$.
                           \begin{itemize}
                             \item For variables of \textbf{Type A} and \textbf{Type B}: skip;
                             \item For variables of \textbf{Type C}:
                             for each edge $(a,i)$, calculate the residual $\gamma_{a\rightarrow i}^t$ using equation $(7)$.
                             Select the edge $(a_\gamma,i_\gamma)$ with the largest residual among all unmarked edges and mark it.
                           \end{itemize}
                \item[3.2] For $i\in V(a_\gamma)$, use equation $(4)$ to calculate $\eta_{i\rightarrow a}^{t}(\sigma_i)$, where $a\in V(i)\backslash a_\gamma$. Update $\mu_{a\rightarrow j}^{t+1}(\sigma_j)$ using equation $(5)$, where $j\in V(a)\backslash i$.
                \item[3.3] Determine whether the iterative equations converge or not.
                           \begin{itemize}
                             \item For variables of \textbf{Type A} and \textbf{Type B}: skip;
                             \item For variables of \textbf{Type C}: for each edge $(a,i)$, if $|\mu_{a\rightarrow i}^{t+1}(\sigma_{i})-\mu_{a\rightarrow i}^{t}(\sigma_i)| <\varepsilon$ holds,
                             break the loop and set $\mu_{a\rightarrow i}^{\ast}(\sigma_i)=\mu_{a\rightarrow i}^t(\sigma_{i})$.
                             Otherwise go to \textbf{3.1} until all edges are marked.
                           \end{itemize}
           \item[3.4] If $t=t_{\rm max}$, output `UNCONVERGED'.
              \end{description}
\end{description}
\rule[1pt]{86.0mm}{1pt}

\section{Results}

\begin{table}[htbp]
\caption{
 \label{tab:real_networks}
For different problem size $n$, the domain size $d_n$, the constraint number $m$, and the theoretical satisfiability threshold $p_s$ obtained by Theorem 1 are shown correspondingly.}
\begin{center}
\begin{tabular}{cccccc}
\hline
{$n$}\hspace{0.8cm}  & {$\alpha$}  \hspace{0.8cm}   & {$r$}  \hspace{0.8cm}   & {$d_n$} \hspace{0.8cm}    & {$m$}  \hspace{0.8cm}    & {$p_s$}\\
\hline
{20} \hspace{0.8cm}  & {0.8}       \hspace{0.8cm}   & {3}    \hspace{0.8cm}   & {11}    \hspace{0.8cm}    & {180}  \hspace{0.8cm}    & $\simeq 0.23$ \\
{40} \hspace{0.8cm}  & {0.8}       \hspace{0.8cm}   & {3}    \hspace{0.8cm}   & {19}    \hspace{0.8cm}    & {443}  \hspace{0.8cm}    & $\simeq 0.23$ \\
{60} \hspace{0.8cm}  & {0.8}       \hspace{0.8cm}   & {3}    \hspace{0.8cm}   & {26}    \hspace{0.8cm}    & {737}  \hspace{0.8cm}    & $\simeq 0.23$ \\
{80} \hspace{0.8cm}  & {0.8}       \hspace{0.8cm}   & {3}    \hspace{0.8cm}   & {33}    \hspace{0.8cm}    & {1052} \hspace{0.8cm}   & $\simeq 0.23$\\
\hline
\end{tabular}
\end{center}
\end{table}
\begin{figure}
\begin{center}
 \includegraphics[width = 0.99 \linewidth]{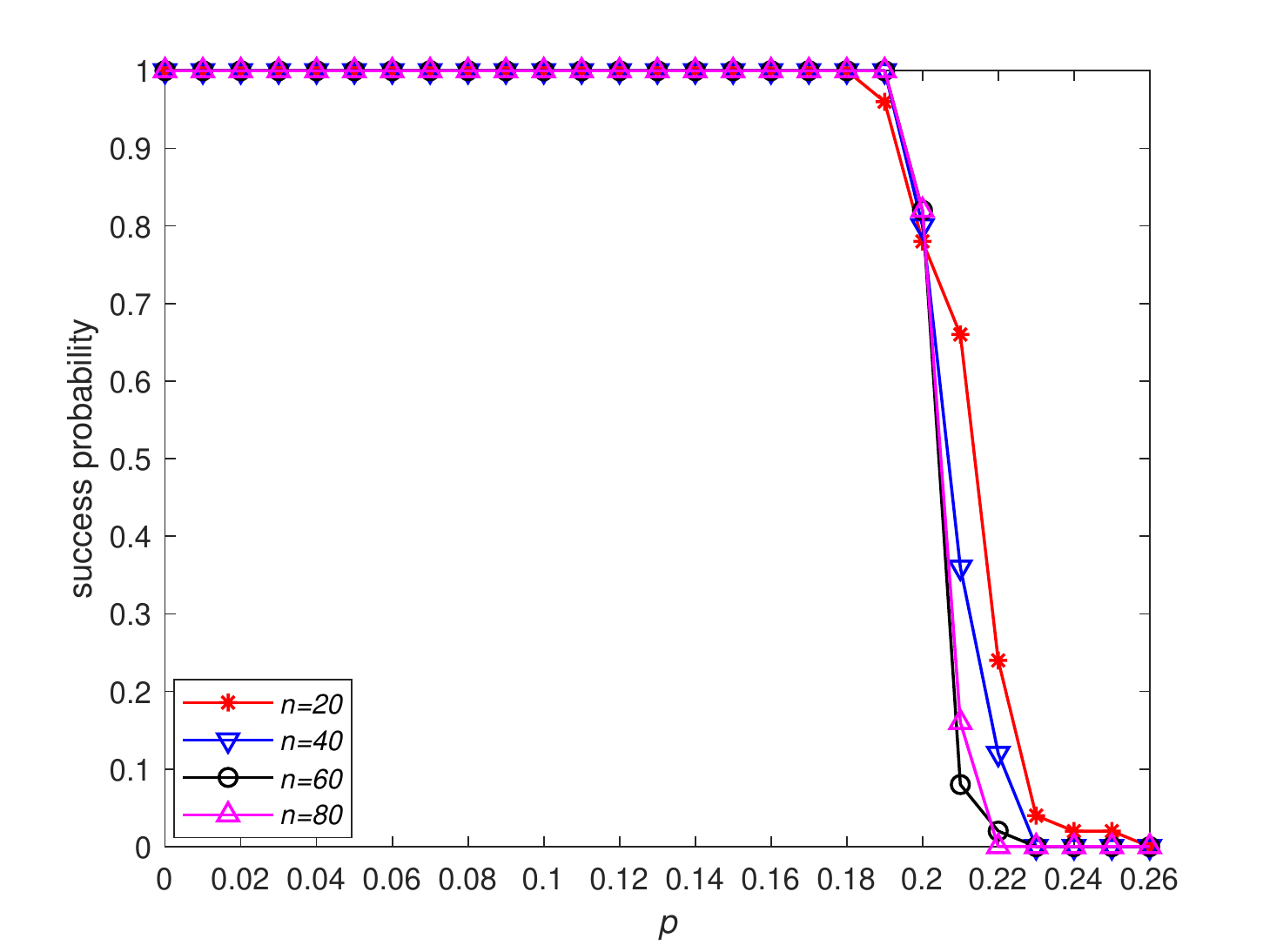} %0.99
\end{center}
\caption{
 \label{fig:fig4}
Fraction of satisfiable instances as a function of $p$ obtained by the MRBP algorithm in solving 50 random instances generated by model RB.}
\end{figure}
\begin{figure}
\begin{center}
 \includegraphics[width = 0.99 \linewidth]{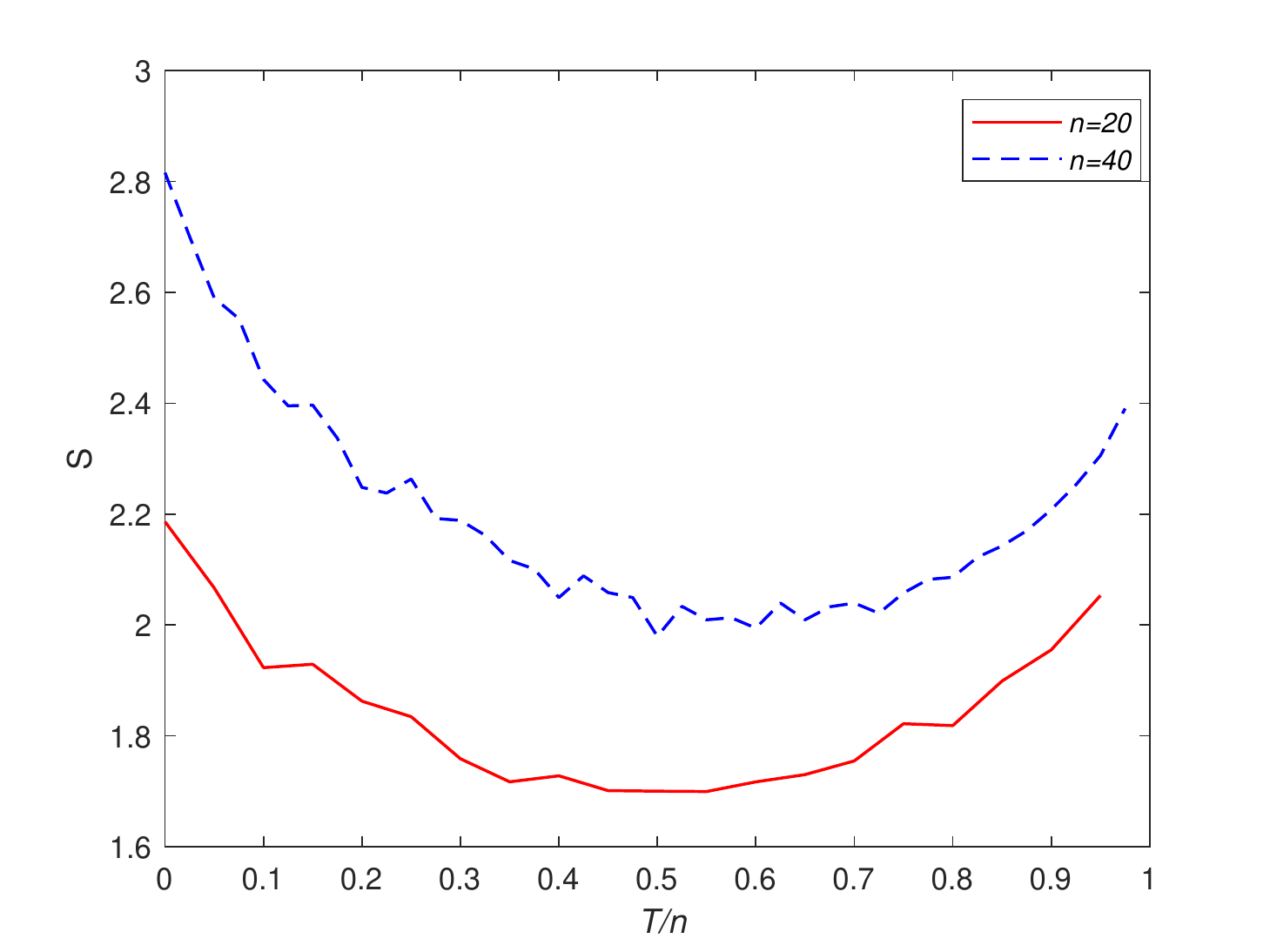} %0.99
\end{center}
\caption{
 \label{fig:fig5}
Entropy $S(x_i^T)$ of the decimated variables at each step $T$  in the MRBP algorithm for binary model RB as a function of $T/n$ at $p=0.1$.}
\end{figure}
\begin{figure}
\begin{center}
 \includegraphics[width = 0.99 \linewidth]{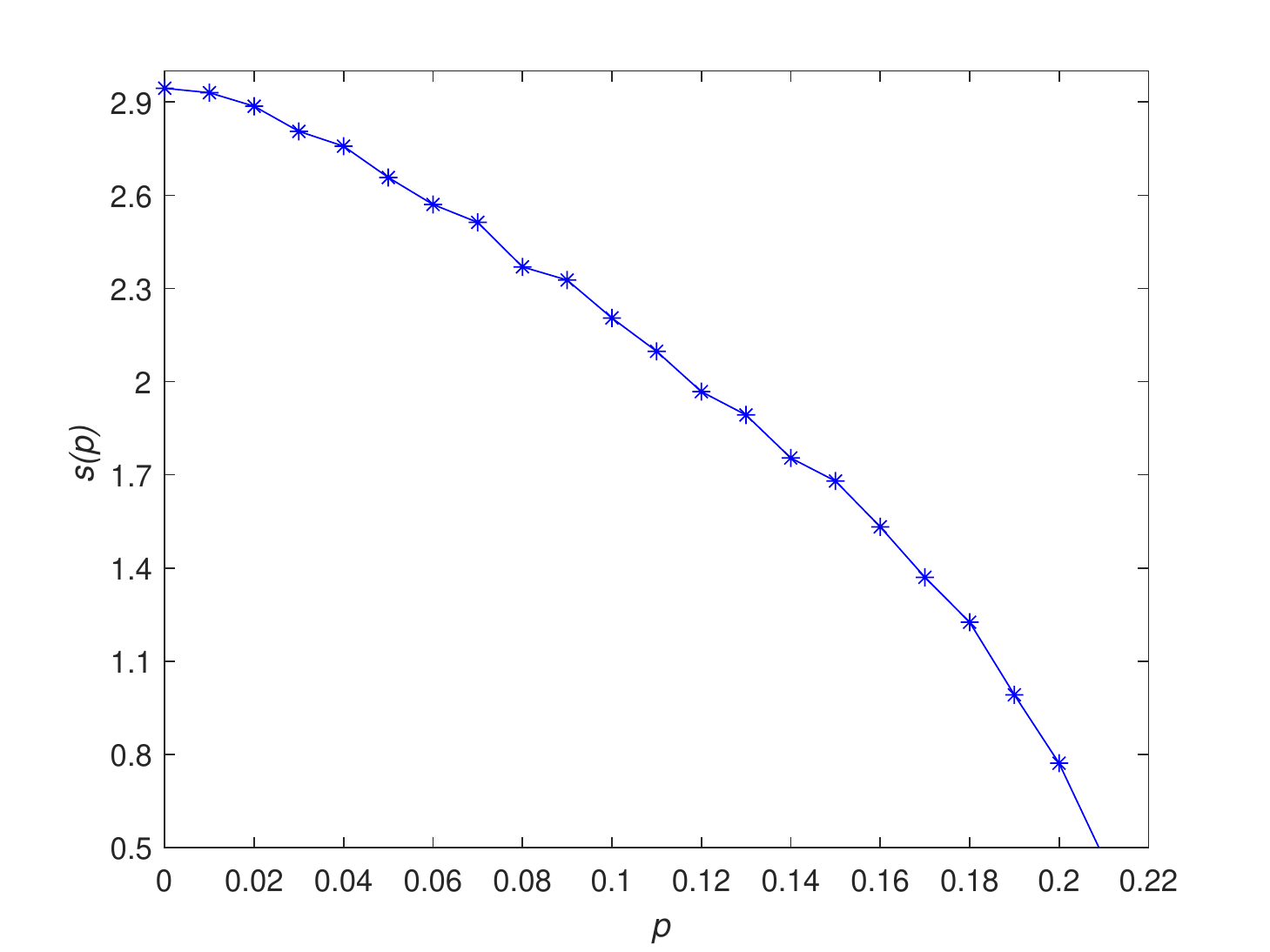} %0.99
\end{center}
\caption{
 \label{fig:fig6}
Mean degree of freedom $s(p)$ on $n$ variables in function of $p$ during the execution of MRBP algorithm on a random instance generated by binary model RB for $n=40$.}
\end{figure}

Here we test our MRBP algorithm based on residuals and show that it not only reduces the fluctuation of the messages, but also significantly improves the convergence of the BP iterative equations. Hence, it can effectively construct a solution of a random instance generated by model RB when approaching the satisfiability threshold.

As a representative parameter set to generate random instances of binary model RB, we take $\alpha=0.8$ and $r=3$. We should mention that, on random instances with other values of parameters of $\alpha$ and $r$, we can obtain similar results on the algorithmic performance around thresholds.  In Table 1, for different problem sizes $n$, corresponding quantities and thresholds are shown. In the MRBP algorithm, we take $t_{\max}=400$ and $\varepsilon=10^{-4}$.

As the first part of our results, we consider the performance of the MRBP algorithm. The fraction of successful runs over $50$ random instances, which are generated by binary model RB for $n\in\{20, 40, 60, 80\}$, is shown in figure 4. It is observed that almost all instances are satisfiable when the constraint tightness $p<0.19$. The MRBP algorithm can construct a solution efficiently when $p<0.20$. However, the algorithm fails with probability tending to 1 when $p>0.22$. Therefore, the algorithm shows a good performance when $p$ approaches the theoretical satisfiability threshold $p_s\simeq 0.23$. Unfortunately, the algorithm fails in the extreme hard region, which may be closely related to the structural transition of the solution space. It is worth noting that the performance of the MRBP algorithm is almost independent of the problem size $n$.

In the decimation process of the MRBP algorithm, we measure the entropy of the selected variable at each time step $T$ for different problem size $n$.
We define the entropy of the decimated variable at $T$ as
\begin{equation}
S(x_i^T) = - \sum_{\sigma_i\in D}\mu_i(\sigma_i)\ln\mu_i(\sigma_i).
\end{equation}
In figure 5, the results of the entropy in function of $T/n$ for $p=0.1$ and $n= \{20, 40\}$ are presented. For a given $n$, the entropy curve shows a concave shape with a minimum at $T/n \approx 0.5$, where half of the variables are fixed with certain values. For $n=20$ and $40$, the entropy $S(x_i^T)$ is always less than the maximum in the case of each variable is evaluated with equal probability from its domain, which guarantees that the MRBP algorithm can select a polarized variable among the remaining free variables to fix.

We then consider the mean degree of freedom of the $n$ variables in the decimation procedure of the BPD algorithm, which is defined as
\begin{equation}
s(p)=\frac{1}{n}\sum_{i=1}^{n}\sum_{\sigma_i\in D} [-\mu_i(\sigma_i)\ln\mu_i(\sigma_i)].
\end{equation}
In figure 6,
the relationship between the mean degree of freedom $s(p)$ and the constraint tightness $p$ for $n=40$ is reported.
It is shown that, with the increase of constraint tightness $p$, $s(p)$ monotonously decreases to $0$. Therefore, we suspect that as $p$ increases, some variables are frozen at certain assignments, which prevents the MRBP algorithm from escaping from the locally optimal solution. Further improving the performance of the algorithm involves studying in more detail on the structural evolution of the solution space of model RB.

\begin{figure*}
\begin{center}
 \includegraphics[width = 0.90 \linewidth]{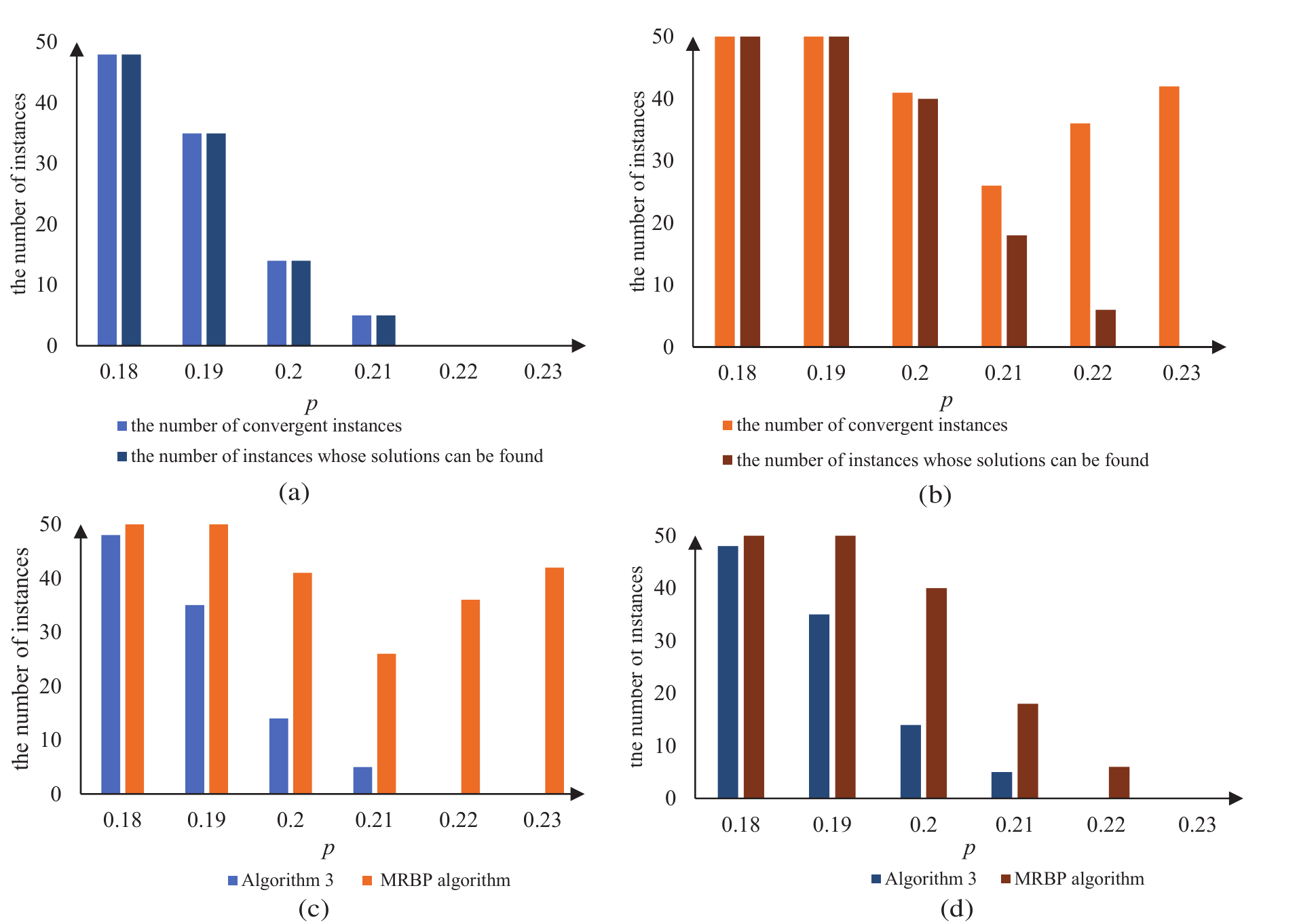} %0.99
\end{center}
\caption{
 \label{fig:fig7}
(a) Number of convergent instances and the number of instances for which a solution can be found by the algorithm 3 in Ref.~\cite{Zhao-JSTAT-2011} when solving $50$ random instances of binary model RB for $n=40$.
(b) Results of the same content with (a) yet obtained from the MRBP algorithm.
(c) and (d) Comparison of the two algorithms on the number of convergent instances and the number of successful instances, respectively.}
\end{figure*}
\begin{figure}
\begin{center}
 \includegraphics[width = 0.99 \linewidth]{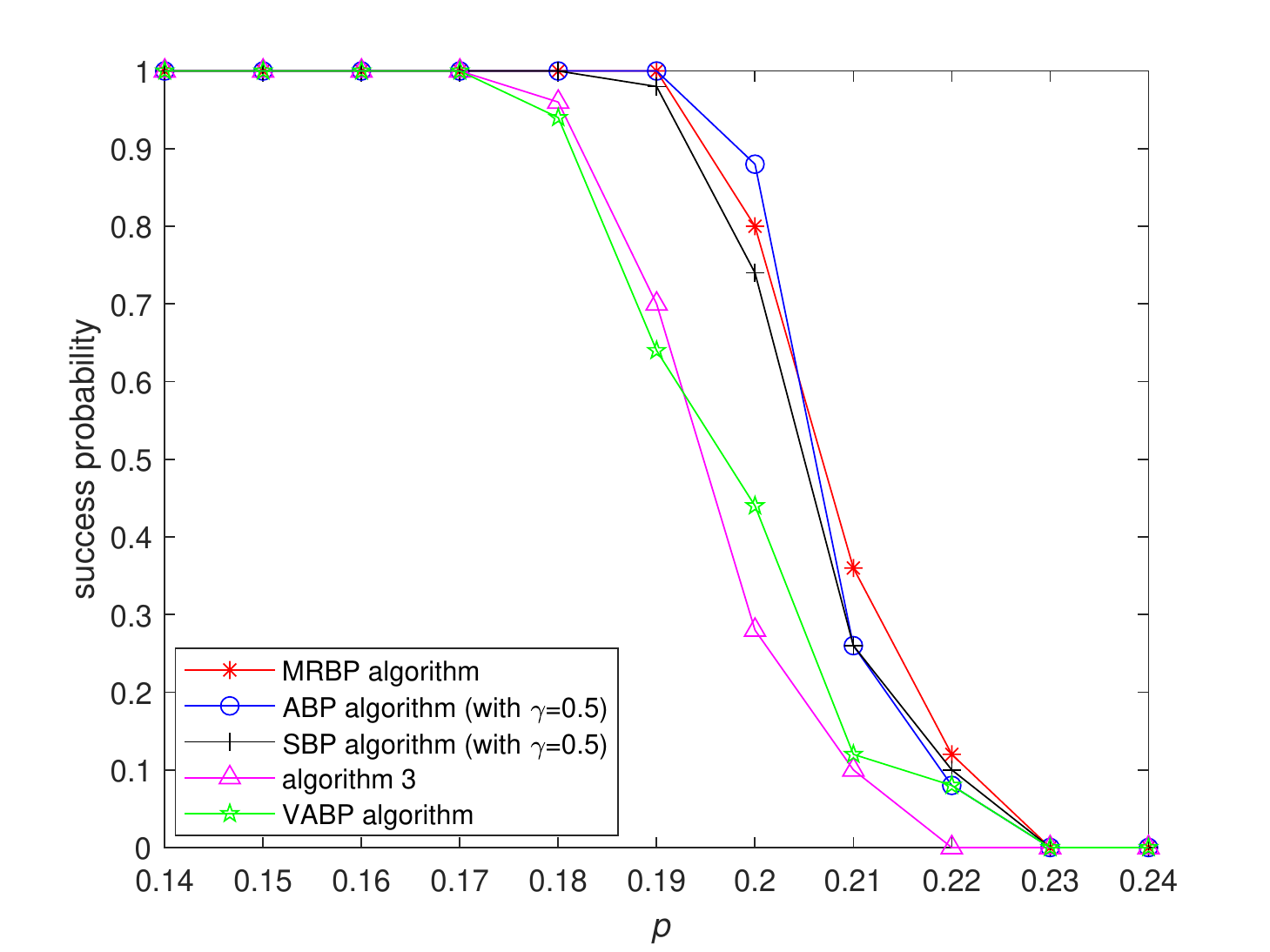} %0.99
\end{center}
\caption{
 \label{fig:fig8}
Comparison of the successful probability among the MRBP algorithm, ABP algorithm with $\gamma=0.5$, SBP algorithm with $\gamma=0.5$, VABP algorithm in Ref.~\cite{Zhao-JSTAT-2021} and algorithm 3 in Ref.~\cite{Zhao-JSTAT-2011}. We consider here $n=40$.}
\end{figure}
\begin{figure}
\begin{center}
 \includegraphics[width = 0.99 \linewidth]{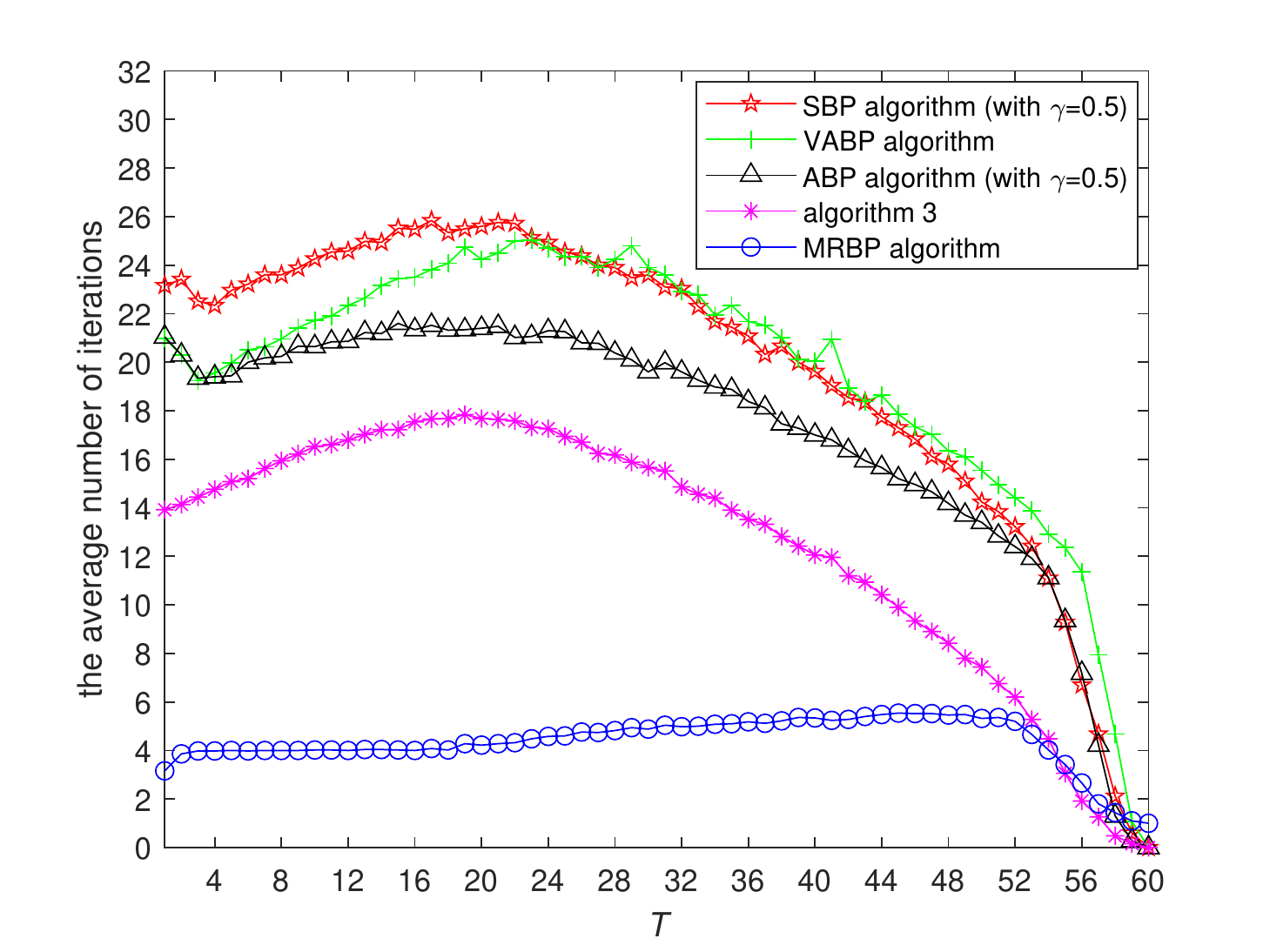} %0.99
\end{center}
\caption{
 \label{fig:fig9}
Comparison of the average number of iterations required by the MRBP algorithm, ABP algorithm with $\gamma=0.5$, SBP algorithm with $\gamma=0.5$, VABP algorithm in Ref.~\cite{Zhao-JSTAT-2021} and algorithm 3 in Ref.~\cite{Zhao-JSTAT-2011} at each decimated step $T$. We consider here $n=60$ and $p=0.15$.}
\end{figure}
\begin{figure}
\begin{center}
 \includegraphics[width = 0.99 \linewidth]{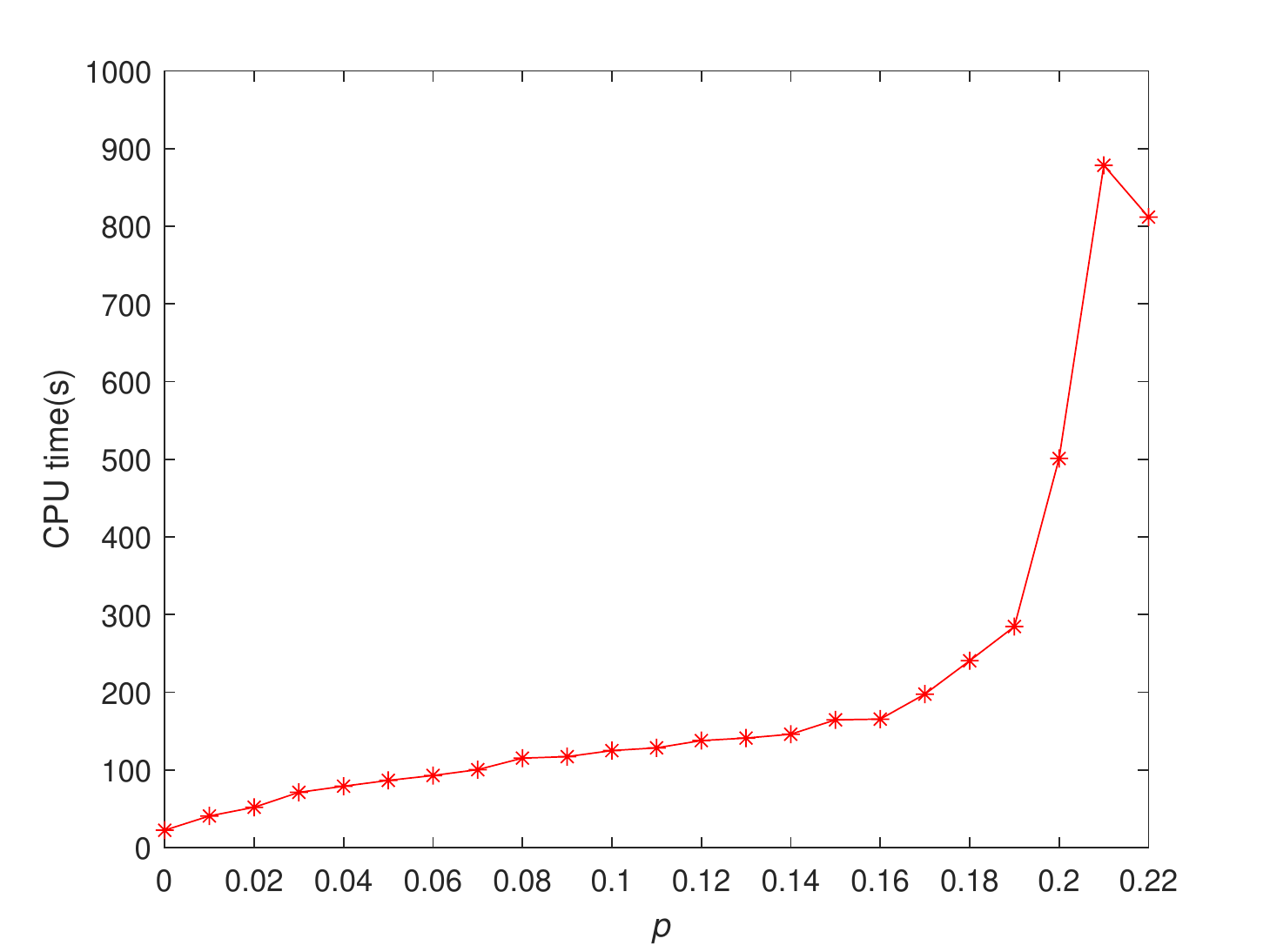} %0.99
\end{center}
\caption{
 \label{fig:fig10}
Running time of the MRBP algorithm in function of $p$. We consider here $n=20$.}
\end{figure}

As the second part of our results, we present a comparison between the usual BP algorithm (algorithm $3$ in Ref.~\cite{Zhao-JSTAT-2011}) and the MRBP algorithm. Figure 7 shows their performance comparison in terms of message convergence and solving power. It can be seen from figure 7 (a), for most instances, if BP algorithm converges, it can usually construct a solution of the instance. However, this is not the case for the MRBP algorithm. As shown in figure 7 (b), when $p< 0.21$, if the MRBP algorithm converges, it almost always finds the solution of the instance. Howerver, when $p\geqslant 0.21$, the assignment obtained after the algorithm has converged may not be the solution of the instance. In other words, compared with the algorithm $3$ in Ref.~\cite{Zhao-JSTAT-2011}, although the convergence of the MRBP algorithm has been greatly improved, the constructed assignment may not be a solution of the instance. As we can see from figure 7 (c), the convergence of the MRBP algorithm is significantly better than the algorithm $3$ in Ref.~\cite{Zhao-JSTAT-2011}. For the MRBP algorithm, the number of convergent instances decreases first and then increases with an increase in $p$. Moreover, the convergence is the worst at $p=0.21$, where the probability of the MRBP algorithm finding a solution of a random instance is extremely low. In figure $7$ (d), the MRBP algorithm shows consistently a higher success probability in finding ground-state solutions. It is obvious that the MRBP algorithm not only improves the convergence of BP iterative equations, but also greatly improves the efficiency of finding solutions.

We also compare the solving efficiency of the MRBP algorithm with three related algorithms based on BP (i.e. ABP algorithm with $\gamma=0.5$, SBP algorithm with $\gamma=0.5$ and VABP algorithm) in Ref.~\cite{Zhao-JSTAT-2021} and algorithm 3 in Ref.~\cite{Zhao-JSTAT-2011}. The results on 50 random instances generated by binary model RB for $n=40$ are shown in figure 8, which illustrate that the MPBP algorithm can find a solution of the problem with a high probability when approaching satisfiability phase transition region.

Furthermore, from the perspective of computational cost, we compare the average number of iterations required at each time step for the decimated variable if BP equations converge. In figure 9, the diagram of average iterations is shown for $n=60$ at $p=0.15$, where the five algorithms can construct a solution of a random instance with probability $1$. We can see that, the number of iterations required by the MRBP algorithm in early decimation process is much lower than that of ABP algorithm with $\gamma=0.5$, SBP algorithm with $\gamma=0.5$, VABP algorithm in Ref.~\cite{Zhao-JSTAT-2021} and algorithm 3 in Ref.~\cite{Zhao-JSTAT-2011}. After approximately $90\%$ of the variables are fixed, the iterations required by the five algorithms are drastically reduced. Therefore, compared with other novel related algorithms, the MRBP algorithm greatly improves the convergence of message passing algorithms.

Finally, we analyze the time complexity of the MRBP algorithm.
The typical time complexity of BP algorithms is $O(M)$ as $M$ is the size of edges in a graph instance.
In our modification on BP algorithms, there are extra $O(1)$ message updating steps after each iteration of messages on all edges.
Thus the complexity of the running time scales as $O(n^2\ln n)$, in which $n^{2} \ln n$ is simply the size of edges in model RB here.
In figure 10, we show the running time of the MRBP algorithm in function of $p$ for $n=20$. We can see that the running time increases sharply when $p$ is close to the satisfiability threshold, a common phenomenon for search algorithms approaching thresholds in CSPs.

Summing the above results together, for the problem of model RB,
the MRBP algorithm greatly improves the convergence of BP iterative equations and shows significantly higher probabilities in finding solutions approaching the satisfiability threshold with a large reduction of iterations in computational cost.

\section{Conclusion}

In this paper, we propose an improved message passing algorithm based on residuals of messages to solve CSPs. The residual of messages is to quantify the fluctuation of messages as a time series between two consecutive time steps. As a modification to the usual message updating process, those messages with the maximal residuals are given higher priority in the updating process. We test our MRBP algorithm on model RB, a random CSP with growing domains, which has an exact satisfiability threshold by previous analytical study. Numerical results show that our algorithm outperforms the BP algorithm with a better convergence of the message updating, a higher probability in constructing a solution for the problem, and a lower computational cost.

To explore the potential implication of residuals on optimization problems and message passing algorithms, several lines of research can be carried out as future work. The first one is to combine our algorithm with a detailed description of the landscape of hard optimization problems to develop better problem-solving algorithms. The second one is to extend our algorithm to combinatorial optimization and CSPs with fixed domains, such as the minimum vertex cover problem and the $k$-SAT problem,
which have a more complex and yet a more refined picture of solution space structure of ground-states configurations.
At the same time, the idea of residuals of messages can be introduced into message passing algorithms in contexts other than sparse graphs, such as the approximated message passing algorithms on dense graphs, cluster variational message passing algorithms on lattice structures, and so on.

\section*{Acknowledgements}

J.-H. Zhao is supported by
Guangdong Major Project of Basic and Applied Basic Research No. 2020B0301030008,
Science and Technology Program of Guangzhou No. 2019050001,
the Chinese Academy of Sciences Grant QYZDJ-SSW-SYS018,
and the National Natural Science Foundation of China (Grant No. 12171479).
C.-Y. Zhao is supported by the National Natural Science Foundation of China (Grant Nos.~11301339 and 11491240108).

\end{document}